\documentclass[showpacs,amsmath,nofootinbib,amssymb, reprint,prd]{revtex4-1}
\pdfoutput=1
\usepackage{mathtools}
\usepackage{hyperref}
\mathtoolsset{showonlyrefs}
\usepackage{psfrag}
\usepackage{array}
\usepackage{amssymb}
\usepackage{amsmath}
\usepackage{amsthm}
\usepackage{graphicx}
\usepackage{epstopdf}
	
\usepackage{epsfig}

\def\pb[#1,#2]{\{#1, #2\}}
\def\deb[#1,#2]{[#1,#2]_{\text{D.B.}}}

\def\Or[#1]{{\text{O}}\left({#1}\right)}
\def\dotl[#1,#2]{\left\langle #1,\, #2 \right\rangle}
\def\dotlb[#1,#2]{\left\langle #1,\, #2 \right\rangle}
\def\dotlm[#1,#2]{\left[ #1,\, #2 \right]}
\def\dotp[#1,#2]{(\vect{#1} \cdot\vect{#2})}
\def\aff[#1,#2]{\hat{#1}(#2)}

\def\n4sym{{\cal N}=4 SYM}
\def\>{\rangle}
\def\<{\langle}
\def\weight[#1,#2,#3]{\{(#1),#2,#3\}}
\def\ads[#1]{$\text{AdS}_{#1}$}

\def\tarelr[#1]{\widetilde{a}^{\text{rel}}_{R#1}}
\def\Oright[#1]{{\cal O}_{R#1}}
\def\Oleft[#1]{{\cal O}_{L#1}}
\def\aleft[#1]{a_{L#1}}
\def\arelr[#1]{a^{\text{rel}}_{R#1}}

\newcommand{\lpl}{\ell_{\text{pl}}}
\newcommand{\mtr}{M^{\text{tr}}}

\newcommand{\be}{\begin{equation}}
\newcommand{\ee}{\end{equation}}
\newcommand{\ba}{\begin{align}}
\newcommand{\ea}{\end{align}}
\newcommand{\bs}{\begin{split}}
\def\sess\end{split}

\newcommand{\vect}[1]{{#1}}

\begin{document}
\title{The Breakdown of String Perturbation Theory for Many External Particles}
\author{Sudip Ghosh}
\email{sudip.ghosh@icts.res.in}
\affiliation{International Centre for Theoretical Sciences, Tata Institute of Fundamental Research, Shivakote, Bengaluru 560089, India.}
\author{Suvrat Raju}
\email{suvrat@icts.res.in}
\affiliation{International Centre for Theoretical Sciences, Tata Institute of Fundamental Research, Shivakote, Bengaluru 560089, India.}
\begin{abstract}
We consider massless string scattering amplitudes in a limit where the number of external particles becomes very large, while the energy of each particle remains small. Using the growth of the volume of the relevant moduli space, and by means of independent numerical evidence, we argue that string perturbation theory breaks down in this limit. We discuss some remarkable implications for the information paradox.
\end{abstract}
\maketitle
\section{Introduction}
Extreme limits of a theory, where perturbation theory breaks down, are often of great interest. In string theory, general arguments that perturbation theory breaks down at high-orders in the genus expansion \cite{Shenker:1990uf,Gross:1988ib} led
to the discovery of D-branes. In this paper, we consider a new, and relatively unexplored, regime of string scattering amplitudes, where the number of external particles, $n$, becomes large, even as the energy per particle remains small and show that a consideration of tree-level scattering already indicates a breakdown of perturbation theory in this limit.

Our analysis ties together seemingly disparate areas of research. We utilize results from the active mathematical literature on volumes of Weil-Petersson moduli spaces, and combine this with extensive numerical analysis of the ``scattering equations'' that have been studied in the literature on amplitudes. We then describe a surprising application of these results to the information paradox: this breakdown in perturbation theory, which may indicate a loss of exact locality, occurs precisely at the point where we expect nonlocal effects to be relevant in black-hole evaporation.

Our results are as follows. We consider the scattering of massless particles in bosonic closed string theory in $d=26$ dimensions in a limit where the dimensionless string coupling constant tends to zero, $4 \pi^2 g_s^2 = \lpl^{d-2} (\alpha'/2)^{1-d/2} \rightarrow 0$, the number of particles $n \rightarrow \infty$, while the energy per particle goes to zero as a power of $n$ so that ${\log(E \sqrt{\alpha'}) \over \log(n)} \rightarrow -\gamma <  0$ with $(d-2) \gamma < 1$.  Then we argue that string perturbation theory breaks down {\em at most} at a value of $n$ that satisfies
\be
\label{largenbreakdown}
{\log(g_s) \over \log(n)} = {(d-2) \gamma - 1 \over 2} + \Or[{1 \over \log(n)}]. 
\ee
We can rephrase this bound directly in terms of the Planck length to read
\be
\label{planckbreakdown}
{\log( E \lpl) \over \log(n)} = {1 \over 2-d} + \Or[{1 \over \log(n)}].
\ee

Our arguments rely on simple unitarity bounds on the growth of tree amplitudes. String amplitudes with $n$ particles may be formulated as integrals over the Weil-Petersson moduli space of Riemann surfaces with $n$ punctures \cite{D'Hoker:1987pr,*Sonoda:1987ra}. In the mathematical literature, these volumes are estimated to grow as $n!$ \cite{mirzakhani2013growth,*zograf1993moduli}. Using bounds on the string integrand, we argue that at large $n$ the growth of the amplitude is controlled by the volume of moduli space, which leads to the result \eqref{largenbreakdown}. 

As independent evidence, we also present a numerical study of high-point string scattering amplitudes. In the large-$n$ limit, we argue that string scattering is dominated by a set of saddle points, just like the high energy limit \cite{Gross:1987kza,*Gross:1987ar}. These saddle point are determined by the ``scattering equations'' \cite{Cachazo:2013gna}, and by solving the scattering equations for a large number of particles, and studying their statistical properties, we verify the $n!$ growth of the tree amplitude. 

The breakdown of perturbation theory is always interesting, but is especially significant in a theory of gravity. This is because {\em locality} is an inherently perturbative notion in quantum gravity. The full quantum gravity path-integral involves a sum over all metrics. Therefore the concept of locality is meaningful only when the S-matrix is calculated in an asymptotic expansion about a given saddle point. The breakdown of perturbation theory indicates that the saddle-point approximation has broken down. Even if the S-matrix can be resummed about a different saddle point (which may not always be possible) the notion of locality in the original saddle is lost. 

This is especially relevant for the information paradox, as we describe in the last section. Most versions of the information paradox in flat space --- such as the ``cloning paradox''\cite{Susskind:1993mu} or the ``strong subadditivity paradox'' \cite{Mathur:2009hf,*Almheiri:2012rt} --- rely on a contradiction that arises if an outside-observer collects a large fraction of the emitted Hawking quanta and then tries to reconcile these observations with properties deduced from {\em locality} in the black-hole spacetime. We show that the breakdown in perturbation theory \eqref{largenbreakdown} kicks in {\em precisely} at this point. This provides a cautionary signal that notions of locality in the original spacetime may be modified. 

This project was motivated by the observation that, in the AdS/CFT correspondence \cite{Maldacena:1997re,*Gubser:1998bc,*Witten:1998qj}, to resolve the information paradox,  it is crucial that  the 1/N expansion breaks down for correlators with $\Or[N]$ insertions \cite{Papadodimas:2012aq,*Papadodimas:2013kwa}. In AdS this breakdown can be tied precisely to the loss of bulk locality in explicit examples \cite{Banerjee:2016mhh}. 

What is the flat-space analogue of the breakdown of the 1/N expansion for high-point AdS correlators? We believe that \eqref{largenbreakdown} provides the answer to this question. 

For convenience, we set $\alpha' = 2$ below.

\section{Bounds on the Growth of Tree Amplitudes}
First, we review some well known bounds on the rate of growth of the tree amplitude in perturbation theory. Unitarity relates the imaginary part of a forward scattering amplitude to the sum of scattering into all possible final states.
\be
\label{unitarityrelation}
\int d \Pi_f {|M(\{k_i\}  \rightarrow \{f\})|^2}  =  2 \text{Im}\left[M(\{k_i\}  \rightarrow \{k_i\}) \right],
\ee
where $d \Pi_f$ is the phase space measure that we specify in more detail below. 

Now, in perturbation theory, the leading terms of \eqref{unitarityrelation} relate the product of tree amplitudes on the left hand side to the imaginary part of loop amplitudes on the right. The validity of perturbation theory requires loop amplitudes to be smaller than tree amplitudes. Therefore, focusing on the case of $n$-point scattering in the center-of-mass frame, and a particular final state with $n/2$ identical  particles, we obtain the following condition for the validity of perturbation theory.
\be
\label{perturbineq}
\begin{split}
&\int  d \Pi_{n \over 2}  |\mtr(k_1 \ldots k_{n\over 2} \rightarrow k_{{n\over2}+1} \ldots k_{n})|^2  \\  &\leq  2 |\mtr(k_1 \ldots k_{n \over 2} \rightarrow k_1 \ldots k_{n \over 2})|.
\end{split}
\ee
The superscript shows that we have only tree amplitudes on both sides of the inequality and
\be
d \Pi_{n \over 2} = {(2 \pi)^d \delta({n E \over 2} - \sum |k_l|) \delta^{d-1}(\sum k_l) \over (n/2)!} \prod_{t} {d^{d-1} k_t \over (2 \pi)^{d-1} 2 |k_t|},
\ee
where $E$ is the center-of-mass energy per particle and $l,t \in \{{n \over 2} + 1, \ldots n\}$. The volume of phase space is given by 
\be
\int d \Pi_{n \over 2} = v {  E^{{(d-2) n \over 2} - d} \over (n/2)! }.
\ee
Here,
\be
v  = {2 \pi \Gamma({d \over 2} - 1)^{n \over 2} (n/2) ^{n({d \over 2} -1) - d} \over (4 \pi)^{(n-2) {d \over 4}} \Gamma\big({(d-2)(n-2)\over4}\big)\Gamma\big({(d-2)n\over4}\big)},
\ee
will be irrelevant below as it is subleading compared to $n!$: $\log(v)/(n \log n) \rightarrow 0$ in the limit under consideration. 

If, at large $n$, tree amplitudes grow as
\be
\mtr(k_1 \ldots k_{n\over 2} \rightarrow k_{{n\over2}+1} \ldots k_{n})= {n! \over \Lambda^{{(d-2) n \over 2} - d}},
\ee
where $\Lambda$ is a characteristic energy scale that must appear on dimensional grounds, then we see that inequality \eqref{perturbineq} is violated at a value of $n$ that satisfies
\be
{(2 - d) \log{E \over \Lambda} \over \log(n)} = 1 + \Or[{1 \over \log(n)}].
\ee

This phenomenon is well known in ordinary quantum field theories. For example, in ordinary $\lambda \phi^4$ theory in four dimensions, $\Lambda = {E/\sqrt{\lambda}}$; so perturbation theory breaks down for $n=\Or[1/\lambda]$  \cite{Libanov:1997nt}. We remind the reader that this scalar theory obeys microcausality nonperturbatively and it is {\em only} in a theory of dynamical gravity that perturbative breakdown indicates a possible loss or change in the notion of locality.

We will now argue that tree amplitudes in string theory also display at least a factorial growth with $\Lambda =  \lpl^{-1}$. The bound \eqref{largenbreakdown} follows immediately.

\section{Growth of String Amplitudes: Analytic Arguments}
First we consider some analytic arguments for this result, which also make contact with some recent mathematical literature.

Usually, in the Polyakov formulation, tree amplitudes are written as integrals over the positions of vertex operators on the complex plane.  But they can also be written as integrals over the moduli space of a $n$-punctured sphere with uniform negative curvature, $-1$ \cite{D'Hoker:1987pr, *Sonoda:1987ra}. 
\be
\label{polyakovamp}
\mtr_n = {\cal N} g_s^{n-2} \int d \mu \, \det(P^{\dagger} P)^{1 \over 2} \det(\Delta)^{-{d \over 2}}  Q_n P_n \bar{P}_n.
\ee
Here $d \mu$ is the Weil-Petersson measure on the $(n-3)$-dimensional moduli space of the $n$-punctured sphere, $\det(P^{\dagger} P)$ is the ghost determinant and $\Delta$ is the scalar Laplacian. ${\cal N}$ is a normalization constant that is irrelevant for now. For massless scattering \cite{D'Hoker:1988ta,Polchinski:1998rq}
\be
\label{nptgreens}
\begin{split}
Q_n &=  e^{- {1 \over 2}  \sum_{i \neq j} k_i \cdot k_j G_{i j} },   \\
P_n &=  {\cal L}\{e^{\sum_{i \neq j} {1 \over 2} \epsilon_i \cdot \epsilon_j \partial_i \partial_j G_{i j}  + k_i \cdot \epsilon_j \partial_j G_{i j} } \}, 
\end{split}
\ee
where ${\cal L}$ means that we extract the part that is linear in each polarization vector $\epsilon_i$,  $G_{i j}$ is the Green's function on the worldsheet and $i,j \in \{1, \ldots n\}$.

The integral \eqref{polyakovamp} suffers from the usual divergences that are present in the Polyakov formulation when two punctures collide. They can be regulated by analytically continuing the kinematic invariants \cite{Berera:1992tm}, complexifying the moduli space \cite{Witten:2013pra} or using string field theory \cite{Sen:2014dqa}. Instead we avoid them by simply cutting off moduli space by restricting the smallest hyperbolic closed geodesic on the worldsheet to have length larger than some $\epsilon$. 

In this truncated moduli space, the determinants that appear in \eqref{polyakovamp} were  bounded in \cite{Gross:1988ib} by relating them to special values of the Selberg zeta function \cite{D'Hoker:1988ta}. 
\be
{\log \text{inf} \{ \det (P^{\dagger} P) \} \over n \log(n)} = {\log \text{sup} \{\det(\Delta) \} \over n \log(n)} = 0+\Or[{1 \over \log (n)}].
\ee

The volume of the Weil-Petersson moduli  space has attracted some recent attention. For genus $g$ surfaces with $n$ punctures, this volume, $V_{g,n}$, grows as $g + n \rightarrow \infty$ so that \cite{mirzakhani2013growth,zograf1993moduli} 
\be
{\log V_{g,n} \over (2 g + n) \log(2 g + n)} = 1 + \Or[{1 \over \log(2 g + n)}].
\ee
We pause to note that, with $n = 0$, the formula above immediately yields the famous $(2 g)!$ growth of string amplitudes at large genus \cite{Shenker:1990uf}.  On the other hand, with $g = 0$ but  $n$ large, the formula also  yields the factorial growth in $n$ that we claimed earlier. 

Turning to the terms in \eqref{nptgreens}, the Green's function itself can be bounded by  $|G_{i j}  \epsilon| \leq C$, where $C$ is an $n$-independent constant \cite{ji1993asymptotic, hempel1988hyperbolic}.  

Since the terms in the exponent of $Q_n$ have alternating signs, both due to the indefinite sign of $k_i \cdot k_j$ and due to the indefinite sign of $G_{i j}$, we expect that this exponent is bounded by $n^{1 - 2 \gamma}$,
\be
{\log|\log Q_n| \over (1 - 2 \gamma) \log(n)} \leq 1 + \Or[{1 \over \log(n)}],
\ee
over at least some fraction of the moduli space. 
This explains why we need $\gamma >  0$ since otherwise this exponent would overwhelm the factorial growth of the moduli-space volume at large $n$. 

If we had been considering tachyon amplitudes, we would have been done at this point. The fact that massless amplitudes can be obtained by cutting open tachyon amplitudes strongly suggests that the former should not grow either much faster or much slower than the latter. But to show this rigorously, we must expand $P_n$ in \eqref{nptgreens} to pick out the relevant linear terms in the polarization vectors.  This expansion yields $\Or[n!]$ terms both in the holomorphic and the anti-holomorphic sector but also comes with a product of $n$-terms involving the derivatives of $G_{i j}$.  We are unable to control this expansion precisely at the moment. So, to complete our argument, we turn to numerical evidence in the next section.

\section{Numerical Analysis of String Scattering Amplitudes}
As explained above, the integral in \eqref{polyakovamp} is divergent and cannot be numerically evaluated directly. However, we do have a large parameter in the problem --- $n$. This allows us to focus on the {\em saddle points} of the integral. These saddle points occur in the interior of the moduli space and a sum over saddle points yields a finite answer. This is the sum that we evaluate numerically and we expect that, at large $n$, it will provide a good approximation to the integral. 

First, we switch to the familiar flat worldsheet metric. In usual holomorphic coordinates, the Green's function is $G_{i j} = -  \ln |z_i - z_j|^2$. Then the saddle points of the integral occur when 
\be
\label{scatteringeqns}
E_i = \sum_j {k_i \cdot k_j \over z_i - z_j} = 0, ~ \forall i.
\ee
These equations were first analyzed to study the high-energy limit of string theory  \cite{Gross:1987kza,*Gross:1987ar} but they have recently attracted attention for their utility in computing amplitudes in quantum field theory. Here, we are  interested in their original application to string scattering.

Remarkably, it turns out that these equations have exactly $(n-3)!$ solutions \cite{Cachazo:2013gna}. Thus, on average, there is precisely one saddle point per unit volume in the Weil-Petersson moduli space!

We approximate the scattering amplitude through a sum over saddle-point contributions,
\be
\mtr_n \approx  (4 \pi g_s)^{n-2} \sum_{\{E_i = 0 \}}{ {\cal G} {\cal J}^{-1}} P_n \bar{P}_n \prod_{i \neq j} |z_{i j}|^{k_i \cdot k_j}.
\ee
Here the ghost contribution is ${\cal G} = \prod_{r \neq s} |z_{r s}|^2$ where $r,s \in \{1,2,3\}$; ${\cal J} = |\det(\partial_p E_q)|$ arises from integrating Gaussian fluctuations about the saddle; $p,q \in \{4 \ldots n\}$, and the normalization and other factors combine into the simple overall constant displayed above. 

Of course, we cannot find all the saddle points, and so we examine a random subset and multiply the mean answer by $(n-3)!$. This is essentially a Monte-Carlo approximation to the full sum. To select a random subset, we start from a random point in the complex plane, and then search for a solution to the equations \eqref{scatteringeqns} using Powell's {\em hybrid} algorithm \cite{powell1970hybrid} as implemented in the GNU scientific library \cite{galassi2009gnu}. Selecting this random origin and flowing down to a root is not quite the same as selecting a random root, but this is the best we are able to do in the absence of further analysis of the basins of attraction of the roots of \eqref{scatteringeqns}.

  In addition, we average the answer obtained over arbitrary initial and final momenta by using a uniform distribution in phase space generated as described in \cite{kleiss1986new}.

To reduce numerical errors, it is convenient to choose polarization vectors so that $\bar{\epsilon}_i = \epsilon_i$. This corresponds to a linear combination of a dilaton and a graviton. 

We do not evaluate $P_n$ exactly, since it has an exponential number of  terms but instead focus on two particular terms,
\be
\label{truncation}
P_n \bar{P}_n \approx |\prod_{i} \sum_{j\neq i} {\epsilon_i \cdot k_j \over z_i - z_j}|^2  + \sum_{\pi} |\prod_{l=1}^{n \over 2} {\epsilon_{\pi_{l1}} \cdot \epsilon_{\pi_{l2}} \over (z_{\pi_{l1}} - z_{\pi_{l2}})^2}|^2,
\ee
where the second term is summed over all possible pairings. This is numerically estimated by averaging over a random subset of possible pairings, and then multiplying by the number of possible pairings. The first term (involving the momenta) is important for $\gamma \approx 0$, while the second term dominates for larger values of $\gamma$. 

To study the statistical properties of the amplitude, we generated $1.15 \times 10^7$ solutions of the scattering equations, comprising 500 random solutions for 500 random points in phase space for each even value of n in the range [10,100].  This computation took about 4000 hours of CPU time.

\begin{figure}
\begin{center}
\includegraphics[width=0.4\textwidth]{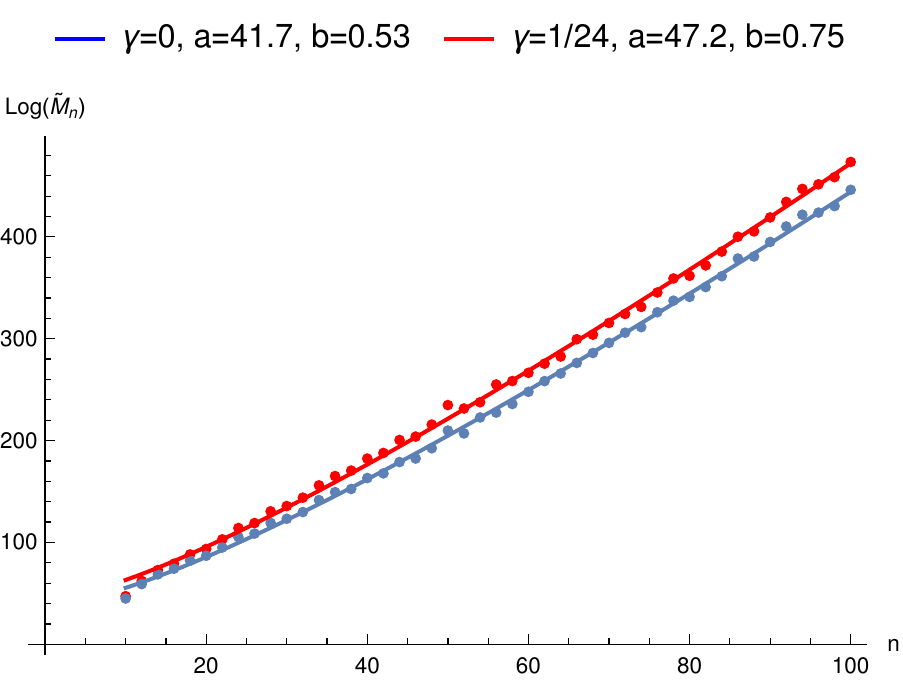}
\caption{Factorial growth of string amplitudes with $n$ \label{numgraph}}
\end{center}
\end{figure}

We show results for two extreme possible values of $\gamma$ in Figure \ref{numgraph}. The data-points show
\be
\log(\tilde{M}_n) = \log(\langle \mtr_n \rangle) - (n-2) \log(4 \pi g_s) +   n \log(d-2),
\ee
 where the last term simply accounts for the average projection of the scattering amplitude on a random set of polarization tensors. The solid curve displays $a + b n  + \log((n-3)!)$, where we set the coefficient of $\log((n-3)!)$ to 1 and use a best-fit for $a$ and $b$. It is clear that this curve fits the data very well and confirms the expected factorial growth of the amplitude.

These numerical results show that our claim of a factorial lower-bound on the growth of the amplitude is robust. But they may underestimate the true growth of the amplitude. This is not only because of the truncation of the prefactor above, but  since we find empirically that the distribution of amplitudes is approximately log-normal. Random sampling may then underestimate the mean.  We discuss this issue further in \cite{sudipsuvratforthcoming}.

\section{Relevance for the Information Paradox}
We now describe the relevance of this breakdown of perturbation theory to the information paradox. First, we recall the cloning and strong-subadditivity paradoxes. A general argument \cite{Page:1993df} tells us that information must exit the black hole after its entropy has reduced by a factor of half. Then, by considering evolution on nice slices that hug the horizon and capture most of the outgoing Hawking radiation, we appear to have a situation where the same information is present in the infalling matter on the slice but also in the exterior. This is the cloning paradox \cite{Susskind:1993mu}. 

It is possible to construct a similar paradox using the strong subadditivity of entropy \cite{Mathur:2009hf,Almheiri:2012rt}. By dividing the black hole into a near-horizon region in the exterior $B$, an analogous interior region $C$, and the rest of the exterior, $A$, we note that after the entropy of the original black hole has shrunk to half its initial value, we have $S_{AB} < S_{A}$; entanglement of modes across the horizon and thermality of Hawking radiation tells us that $S_{B C} < S_C$ and this seems to contradict the strong subadditivity of entropy: $S_{AB} + S_{BC} \geq S_{A} + S_{C}$ \cite{Lieb:1973cp,*lieb1973prl}.  

However, both these paradoxes can be resolved, while preserving an empty interior, by recognizing that the Hilbert space does {\em not} factorize into the space outside and the space inside. Instead, as black hole complementarity \cite{'tHooft:1984re,*Susskind:1993if,*Susskind:1993if} suggests, the degrees of freedom inside the black hole are ``scrambled'' versions of degrees of freedom outside. (For alternate resolutions, see \cite{Mathur:2008kg,Bena:2015dpt}.)

In AdS/CFT, complementarity appears manifestly in the construction of the interior described in  \cite{Papadodimas:2013wnh,*Papadodimas:2013jku,*Papadodimas:2015xma,*Papadodimas:2015jra,*Raju:2016vsu}. The claim is that a field operator inside the horizon can be rewritten as a polynomial of degree $S=\Or[N]$ in field variables outside the horizon: $\phi(x_{\text{in}}) = {\cal P}(\phi(x^{\text{out}}_1) \ldots \phi(x^{\text{out}}_S))$. A similar phenomenon can be seen even in empty AdS where a version of the formula above was derived explicitly in \cite{Banerjee:2016mhh}.  Such a relation is possible because bulk AdS locality breaks down entirely for $N$-point correlators.

We now show how the breakdown of perturbation theory described above makes a similar relation plausible in flat space.  Recall that in $d$-dimensions, the temperature and entropy of a Schwarzschild black hole of radius $R$ are given by
\be
4 \pi T_H = (d-3) R^{-1} , \quad S =  2 \pi (R \lpl^{-1})^{d-2} \Omega_{d-2}.
\ee
Any experiment that involves a significant fraction of the emitted Hawking quanta requires us to measure some connected $S$-point correlators, where each particle has energy of order $T_H$. But, substituting $E = T_H$ in the formula \eqref{planckbreakdown}, we find that {\em precisely} at $n = S$, perturbation theory breaks down! It is possible to repeat the analysis above when some of the extra dimensions are compactified and, once again, perturbation theory breaks down precisely at $n=S$ \cite{sudipsuvratforthcoming}.

So it may be misleading to apply any consequence of causality derived in the original black-hole saddle point to an observation that involves S-point correlators of particles with characteristic Hawking energy. This was done to derive both paradoxes above and our analysis indicates (although it does not prove) that such correlators may receive contributions from an entirely different saddle point with a distinct causal structure.

We emphasize that it is perfectly possible to have large nonlocal effects in very high-point correlators as indicated above, while preserving locality and, in particular, a very small commutator between spacelike-separated operators in low-point correlators.

\section{Discussion}
Beyond implications for the information paradox, this limit of high-point string scattering amplitudes may be useful in other ways.  While the literature has largely focused on the breakdown of perturbation theory at high genus,  the large-$n$ breakdown is more tractable because it can be seen at tree-level, where we can analyze the amplitude explicitly. For instance if, in some cases, it turns out to be possible to resum perturbation theory about another saddle point, it may be easier to examine this resummation in the large-$n$ limit.

We would like to extend this analysis to other settings. The breakdown that we have described above also occurs in quantum field theory (although in that case we would have no restriction that $\gamma >  0$) and what we have shown here is that it cannot be cured by perturbative bosonic string theory.
We expect that the same breakdown should extend to superstring theory in $d=10$ \cite{sudipsuvratforthcoming}. 

In the analysis above, we considered a single channel of ${n \over 2} \rightarrow {n \over 2}$ scattering. The initial particles could also scatter into excited string states or a different number of particles. Although the number of such states grows exponentially as $\exp(n E) \sim \exp(n^{1- \gamma})$, this is still subleading compared to the factorial growth of the amplitude.  So, including these channels may not alter \eqref{largenbreakdown} to leading order but it would be interesting to examine this further.

The problem of $2 \rightarrow n$ scattering, was studied in \cite{Dvali:2014ila, *Addazi:2016ksu}. Our kinematical regime is somewhat different, because we divide the energy democratically between the particles rather than taking two of them to be ultra-Planckian.  Our techniques are also distinct, since we utilize the volume of moduli space and a numerical analysis of the average magnitude of the amplitude in phase space rather than its exact value at special kinematic configurations. However, our results for the scaling of the amplitude appear to be consistent with those of \cite{Dvali:2014ila, *Addazi:2016ksu}, when we extrapolate from their regime to ours. 

It would also be very interesting to extend the analysis of \cite{Banerjee:2016mhh} and find an explicit formula relating the exterior and the interior of a black hole in flat space --- the existence of which we have made plausible in this paper.

\paragraph*{\bf Acknowledgments:}
We are grateful to Rajesh Gopakumar, Subhojoy Gupta, R. Loganayagam, Gautam Mandal, Shiraz Minwalla, Kyriakos Papadodimas,   Ashoke Sen, Sandip Trivedi and also the participants of the Bangalore Area Strings Meeting,  ICTS/Prog-BASM/2016/07, for useful discussions. We are grateful to Ashoke Sen for comments on a preliminary version of this manuscript. We are grateful to TIFR (Mumbai) and the Nordita program on Black Holes and Emergent Spacetime for hospitality, while this work was in progress.  Our numerical analysis was performed using the computing facilities at ICTS; in particular, we acknowledge the use of the high-performance computing cluster {\em Mowgli.} We also acknowledge the use of GNU parallel \cite{Tange2011a}. 
\bibliography{references}
\end{document}